\documentclass[superscriptaddress,twocolumn]{revtex4-1}
\usepackage{graphicx}                   
\usepackage{hyperref}                   
\usepackage{amsmath,amssymb}            
\usepackage{multirow}	
\usepackage{amsmath}
\usepackage{amssymb}	
\usepackage{xcolor}	
\usepackage{amsmath,color}

\graphicspath{{figs/}}					

\DeclareMathAlphabet{\mathpzc}{OT1}{pzc}{m}{it}

\begin{document}
	
	\author{Hediye Yarahmadi}
	\affiliation{Department of Physics, University of Tehran, P. O. Box 14395-547, Tehran, Iran}

	\author{Abbas Ali Saberi}\email{(corresponding author) ab.saberi@ut.ac.ir}
	\affiliation{Department of Physics, University of Tehran, P. O. Box 14395-547, Tehran, Iran}
	\affiliation{Institut f\"ur Theoretische
		Physik, Universit\"at zu K\"oln, Z\"ulpicher Str. 77, 50937 K\"oln,
		Germany}
	
	\title{A 2D L\'evy-flight model for the complex dynamics of real-life financial markets }

\begin{abstract}
We report on the emergence of scaling laws in the temporal evolution of the daily closing values of the S\&P 500 index prices and its modeling based on the L\'evy flights in two dimensions (2D). The efficacy of our proposed model is verified and validated by using the extreme value statistics in random matrix theory.
We find that the random evolution of each pair of stocks in a 2D price space is a scale-invariant complex trajectory whose tortuosity is governed by a $2/3$ geometric law between the gyration radius $R_g(t)$ and the total length $\ell(t)$ of the path, i.e., $R_g(t)\sim\ell(t)^{2/3}$. 
We construct a Wishart matrix containing all stocks up to a specific variable period and look at its spectral properties over 30 years. In contrast to the standard random matrix theory, we find that the distribution of eigenvalues has a power-law tail with a decreasing exponent over time---a quantitative indicator of the temporal correlations. We find that the time evolution of the distance of a 2D L\'evy flights with index $\alpha=3/2$ from origin generates the same empirical spectral properties. The statistics of the largest eigenvalues of the model and the observations are in perfect agreement. 
\end{abstract}
\maketitle

\textbf{Extreme financial events \cite{christoffersen1998horizon, longin2016extreme} are much more common than the ordinary theory of random walks with normal fluctuations anticipates. The financial returns exhibit heavy-tailed distributions \cite{rachev2003handbook} that, in relation to chaos theory, introduce L\'evy stable functions as possible explanations. Here, we report on developing a model based on L\'evy flights in two dimensions whose time-evolving distance from origin generates data with a number of key common characteristics with the recorded daily prices in S\&P 500 stocks. We find a $2/3$ law, akin to the same law in turbulence \cite{mantegna1999introduction}, which describes the complexity of random trajectories traveled by every pair of stocks in the price space as a part of the whole dynamical system. We construct random matrices of (log-)returns over a variable epoch size and show that the eigenvalue spectrum has a power-law behavior with a scaling exponent that decreases with the interval recording observation. The results of our analysis for our proposed model are in perfect agreement with the empirical data from S\&P 500 stocks at every given time interval, unraveling the nature of complex cross-correlations. We believe that our model can serve as a valuable tool to predict risk estimations with the possible assessment of finite sampling interval effects in real-world financial markets. }

\textbf{Introduction.} Deep understanding of interacting complex systems has become an underlying issue in a broad spectrum of interdisciplinary research in diverse fields of condensed matter physics, medicine, psychology, sociology, biology, and computational social
sciences \cite{strogatz2018nonlinear}. A fundamental principle of any complex system is the interplay of nonlinear interactions between the system's components. Economic time series also depend on the evolution of a large number of interacting ingredients, and so are a striking example of such complex evolving systems \cite{liu1999statistical,stanley1995zipf,vemuri2014modeling}. These make economic systems extremely attractive for physicists interested in deeper understanding from the statistical behavior of the financial markets \cite{mantegna1995scaling,kertesz1999econophysics,liu1997correlations,elton2009modern, takayasu1997stable,egenter1999finite,lee1998universal,plerou1999scaling,stanley1996scaling,laloux1999noise,amaral1998power,mantegna1999hierarchical,plerou1999similarities,ghashghaie1996turbulent,mantegna1996turbulence,bouchaud2001leverage,gopikrishnan1999scaling,qiu2006return,stanley2000introduction,plerou2002quantifying,takayasu2006practical,lillo2003master,onnela2003dynamics,kiyono2006criticality,gabaix2003theory,mantegna1999introduction,bouchaud2000theory}, analyzing the correlations  between different stocks and quantifying these correlations \cite{laloux1999noise,liu1997correlations,mandelbrot1997variation,pharasi2018identifying,mantegna1999introduction,bouchaud2003theory}. 

One of the major questions is how to model the rare events lying outside the range of available observations. In such cases, it is necessary to rely on a completely fundamental method. Extreme Value Theory (EVT) is an active field of research in statistical science providing a well-developed tool to model the extremes at tails of distributions of uncorrelated random variables \cite{ majumdar2020extreme,fisher1928limiting,leadbetter2012extremes,gumbel1958statistics,fortin2015applications}. The limiting distribution of the extremes exhibits some degree of universality depending on the microscopic distribution: for distributions that vanish beyond a finite value one finds Weibull, for distributions decaying faster than any power-law (like an exponential distribution) one finds Gumbel, and power-law distributions lead to the Fr\'echet distribution \cite{ majumdar2020extreme,fisher1928limiting,leadbetter2012extremes,gumbel1958statistics,fortin2015applications}. Recently, there have been made some advances in understanding of EVT of correlated variables as well \cite{majumdar2020extreme}. EVT can also be applied to forecast crashes and extreme loss situations. Extraordinary performance of EVT in tail modeling makes it a beneficial tool in risk-related topics \cite{aslanertik2017extreme,gilli2006application,bensalah2000steps,loretan1994testing,longin1996asymptotic,danielsson2000value, mcneil2000estimation,jondeau1999tail,neftci2000value,genccay2003high}. In the context of extreme price movements of the financial stock market, it is shown that the distribution of the lowest daily return and the highest daily return of the stock market index can be given by the Fr\'echet distribution \cite{longin1996asymptotic,mcneil2015quantitative,gangwal2018extreme}.

An appealing solvable example of the extremal statistics of strongly correlated variables is Random Matrix Theory (RMT) concerning the distribution of the bulk and edge eigenvalues \cite{tracy1994level,tracy1996orthogonal,majumdar2014top, dean2006large,dean2008extreme}. It helps to illuminate the difference between random and non-random information \cite{laloux1999noise,plerou1999universal,gopikrishnan2001quantifying,plerou2002random}.\\
RMT has been applied extensively in the investigation of time series of financial markets and is one of the immensely used methods for studying the correlations in stocks \cite{mantegna1999information,mantegna1999hierarchical,laloux1999noise,shen2009cross,plerou1999universal,gopikrishnan2001quantifying,kullmann2002time,onnela2003dynamics,plerou2002random,pharasi2018identifying,pharasi2019complex,paul2014random,utsugi2004random,akemann2010universal,mantegna1999introduction,bouchaud2000theory,kumar2012correlation,wang2013random,gopikrishnan1999scaling, namaki2020analysis}. Analyzing the properties of the cross-correlation matrix (\textsf{C}) on several stock markets was demonstrated to agree with RMT predictions whose elements are uncorrelated \cite{laloux1999noise,plerou1999universal,plerou2002random}.  Agreement of the eigenvalue statistics of \textsf{C} with RMT results implies that \textsf{C} has entries that contain a significant degree of randomness \cite{plerou2000random}. Also, the statistics of eigenvalues and the largest eigenvalues are found to follow the semicircle and the Gaussian Orthogonal Ensemble (GOE), respectively \cite{plerou1999universal}. The analysis of eigenvalues that deviate from RMT shows the existence of cross-correlations between stocks \cite{gopikrishnan2001quantifying,mantegna1999hierarchical}.

\begin{figure}
	\centering
	\includegraphics[width=3.4in]{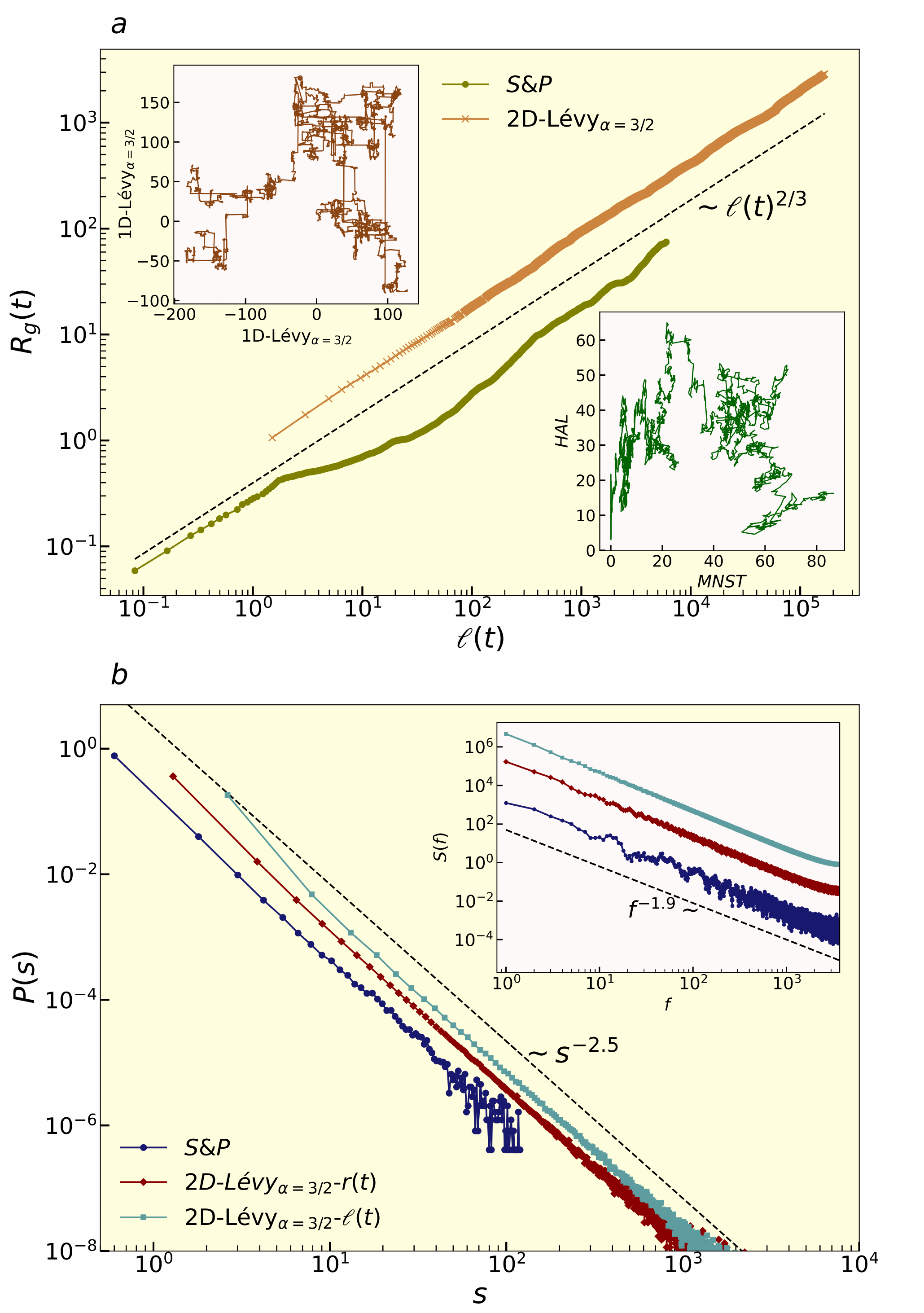}
	\caption{(a) Main: The average radius of gyration $R_g(t)$ vs the average total length $\ell(t)$ of a pair-stocks trajectory both for S\&P and 2D-L\'evy$_{\alpha=3/2}$ giving rise to the same fractal dimension $d_f=3/2$. Inset: Sample geometries are shown for a pair of stocks in their 2D price space (bottom-right) and two independent 1D-L\'evy flights (top-left).
	(b) Main: The distribution $P(s)$ of the price increments $s$ for S\&P, compared with the distribution of the step lengths of 2D-L\'evy$_{\alpha=1.5}$-$r(t)$ and $\ell(t)$. All are consistent with a power-law $\sim s^{-(1+\alpha)}$ with $\alpha=1.5$. Inset: The power-spectrum of the three sets shown in the Main panel exhibit long-range correlations with the same power exponent $\beta\simeq 1.9$. The total number of time steps for S\&P data is $7740$, and for the L\'evy flights is $10^5$ in the Main panel (a) and $7740$ in (b) and in the whole text. We used an ensemble of $262\times100$ independent 2D-L\'evy samples for our averaging in our study.
}
	\label{fig1}
\end{figure}

Various models and theoretical approaches have been developed to interpret the features of the financial dynamics \cite{shen2009cross,eguiluz2000transmission,krawiecki2002volatility,lux1999scaling,zheng2004two,ren2006minority,kumar2012correlation,hosseiny2016metastable, bahrami2020optimization, shirazi2017non, hosseiny2019hysteresis, hosseiny2017geometrical}, for example in \cite{peron2011collective,mantegna1999hierarchical,vandewalle2001non,onnela2003asset,onnela2003dynamic} financial markets have been considered by using concepts from complex networks theory. Financial markets have been also modeled by continuous diffusion processes \cite{bachelier1900theorie}, such as Brownian motion, and discontinuous processes \cite{mandelbrot1997variation}, like L\'evy processes. In \cite{devi2021asymmetric, devi2017financial}, the long-term behavior of the financial market returns is investigated.  The results indicate that both superstatistics and Tsallis statistics non-extensive models have to be considered to describe the complex dynamics of financial markets.\\
In the present work, our purpose is to introduce a parameter-free model that can ideally describe financial markets such as S\&P stocks. Our analysis singles out a specific class of the family of L\'evy flights in two dimensions which models the dynamics of S\&P returns. The privilege of this L\'evy flight model is that it can generate the statistical properties of S\&P data without tuning any additional parameters and can act as a valuable tool to predict risk estimations and assess finite sampling interval effects in real-world financial markets.

\textbf{Data and theory.} We study the structure and dynamics of the stock market S\&P 500 index, or the Standard \& Poor's $500$ index from the New York Stock Exchange (NYSE) containing the "adjusted closing prices" of $262$ stocks at one-day intervals traded from the period January $2$, $1990$ to September $18$, $2020$, extracted from 
\href{http://finance.yahoo.com}{finance.yahoo.com}.\\
We calculate the logarithmic increments \begin{equation}\label{log-return}
\delta R^{\prime}_i(t)\equiv\ln P_i(t+1)-\ln P_i(t),
\end{equation} 
where $P_i(t)$ denotes a price at time $t$ of the $i$th stock $(i = 1, 2, ..., N=262)$ and the time $t$ runs over the $30$-years period $1990$–$2020$ (with the total number of $7740$ trading days). The reason for analyzing the returns $\delta R^{\prime}_i(t)$ rather than the actual raw asset prices $P_i(t)$ is that it gives a scale-free assessment of the performance of the asset with attractive statistical properties. Since different stocks have varying levels of volatility (standard deviation), we define a normalized return 
\begin{equation}\label{normal-return}
	\delta R_i(t)\equiv\frac{\delta R^{\prime}_i(t)-\left<\delta R^{\prime}(t)\right>}{\sigma(t)},
\end{equation}
where $\left<\cdots\right>$ denotes the ensamble average at time $t$, and
$\sigma(t)=[\left<\delta R^{\prime}(t)^2\right>-\left<\delta R^{\prime}(t)\right>^2]^{1/2}$. There is a considerable interest in the financial literature in the recognition of the log-return densities. 
It has been confirmed that log-return distributions of financial indices reveal heavier tails and are more peaked than the Gaussian assumption would permit. Actually, the student's t-distribution with $3.0$-$4.5$ degrees of freedom was identified as the best fit to daily log-returns \cite{markowitz1996likelihood,markowitz1996likelihood,hurst1997marginal,fergusson2006distributional,praetz1972distribution,blattberg2010comparison,mcneil2015quantitative,platen2008empirical}.

In order to devise a model that can ideally describe such a financial market, one may think of correlated random walks \cite{gillis1955correlated}, or Brownian motion \cite{uhlenbeck1930theory,wang1945theory} with Gaussian statistics that suppresses large jumps. However, the observed t-distribution signals the existence of correlations and the possibility of large jumps. These resulted in a very special case of random walks with heavy-tail jump distributions called L\'evy flights \cite{mantegna1995scaling,mantegna1994stochastic,shlesinger1995levy,yang2010engineering,guyon1993anomalous,metzler2000random,montroll1984nonequilibrium,hughes1981random} to study the extensive presence of large fluctuations in econophysics \cite{mantegna1995scaling} and human neuroscience \cite{cabrera2004human}. There has been also developed a theoretical model of the truncated L\'evy flight that describes several statistical features of the S\&P stock index \cite{mantegna1995scaling,mantegna1994stochastic,mantegna1999introduction}.

L\'evy flight is a Markovian stochastic process whose step length $s$ obeys the power-law distribution 
\begin{equation}\label{power-dist}
	P(s)\sim \left|s\right|^{-(1+\alpha)},
\end{equation} for large $s$ with $0<\alpha< 2$. Due to the divergence of their variance $\langle s^2(t)\rangle\rightarrow\infty$, it allows for the occurrence of extremely long jumps. The trajectory of a L\'evy flight is a self-similar  object \cite{yang2010engineering, hughes1995random} with fractal dimension 
\begin{equation}\label{d_f}
d_f=\alpha.
\end{equation}
\textbf{Scale invariance of evolving geometry.} We hypothesize that it may be possible to model the dynamics of S\&P prices by the L\'evy flight in two dimensions (the numerical details to generate the time-series of 2D L\'evy-flight can be found in the Supplementary Sec. I). To this aim, we have to first determine the most suitable value for $\alpha$ which agrees with the real data. Therefore, we use the relation (\ref{d_f}) to best estimate $\alpha$ by measuring the fractal dimension of a trajectory that a pair of stocks travel in their 2D price space (for instance, see a sample trajectory shown in the bottom-right Inset in Fig. \ref{fig1}(a) for two randomly chosen stocks in S\&P).  Figure \ref{fig1}(a) shows the average gyration radius $R_g(t)=\big\langle\big[\frac{1}{t}\sum_{k=1}^{t}r^2_k\big]^{1/2}\big\rangle$ versus the average total length $\ell(t)$ of such trajectories ($r_k$ denotes the distance from origin at the $k$th time step (see the Supplementary Sec. II for more details). The averages are taken over all trajectories produced by every two different stocks in S\&P. We examine the following scaling relation 
\begin{equation}\label{d_f-R_g}
R_g(t)\sim \ell^{1/d_f}.
\end{equation}
Our best estimate offers a good agreement with Eq. (\ref{d_f-R_g}) with $d_f=3/2$ (Fig. \ref{fig1}(a)). According to Eq. (\ref{d_f}), this fractal dimension also holds for the trajectory of a 2D L\'evy flight with $\alpha=3/2$ (a putative trajectory is shown in the top-left Inset in Fig. \ref{fig1}(a) which is comparable with that shown for a pair of stocks in S\&P in the bottom Inset). We have also shown the relationship (\ref{d_f-R_g}) for an ensemble of 2D-L\'evy$_{\alpha=3/2}$ trajectories which is in perfect agreement with the theoretical prediction shown by the dashed line in Fig. \ref{fig1}(a).

One of the main objectives of this paper is to be able to use this 2D-L\'evy$_{\alpha=3/2}$ model to predict the temporal evolution of prices in the S\&P market, so that we can see the most similarity between the statistical properties of time series in real markets and the predictive model. To this end, we noticed that the generation of this time series can be done in two ways: first, each price can be considered at any time $t$ equal to the total length $\ell(t)$ traveled by the 2D-L\'evy flight up to that time, or as a second option, the price can be considered at each time $t$ equal to the distance $r(t)$ of the 2D-L\'evy flight at that time from the origin (i.e., the position at $t=0$). We have used $262\times100$ independent samples for both models in our statistical analysis. In order to distinguish these two definitions in the rest of the paper, we will refer to them with 2D-L\'evy$_{\alpha=3/2}$-$r(t)$ and 2D-L\'evy$_{\alpha=3/2}$-$\ell(t)$, respectively.

On the other hand, in order to further support the validity of $\alpha=3/2$ estimated from the geometrical correspondence between the S\&P markets and the 2D-L\'evy$_{\alpha=3/2}$, let us now examine the power-law relation (\ref{power-dist}) for the distribution of the price increments $s$ for every stock contributing in the S\&P over the 30 years. As shown in Fig. \ref{fig1}(b) the distribution of the steps in both definitions of L\'evy model exhibits similar scaling behavior $\sim s^{-(1+\alpha)}$ with $\alpha=3/2$ in agreement with the distribution of the real-life price increments. However, as we will see in the following, more detailed analysis shows that the model based on 2D-L\'evy$_{\alpha=3/2}$-$r(t)$ is statically much more consistent with real-life markets.

\textbf{Extent of correlations.} A universally used method for investigating long-range correlation properties of prices in time series is the power-spectrum analysis \cite{liu1999statistical}. The power-spectrum $S(f)\sim f^{-\beta}$ of a wide-sense stationary random process is the Fourier transform of its autocorrelation function. If the data are uncorrelated, one finds $ \beta = 0 $, while for correlated data the spectral density will be large at small frequencies and small at high frequencies giving rise to a nonzero power exponent $\beta\ne0$. The Inset of Fig. \ref{fig1}(b) shows the average power-spectrum of prices in the S\&P index  measured in a one-day interval using data recorded during the $30$ years. We find that the correlations can be described by a power-law with $\beta\simeq1.9$.  
This is an indication of a long-range correlations in the data. Similar power-spectrum analysis for the time series generated for 2D-L\'evy$_{\alpha=3/2}$-$r(t)$ and 2D-L\'evy$_{\alpha=3/2}$-$\ell(t)$ gives the same exponent $\beta\simeq1.9$ (see the Inset of Fig. \ref{fig1}(b)). 

We have also analyzed the distribution of normal returns of S\&P and 2D-L\'evy$_{\alpha=3/2}$-$r(t)$, $\ell(t)$ as defined in (\ref{normal-return}). As shown in the Supplementary Fig. S2, the estimated distributions of the log-returns of the S\&P and 2D-L\'evy$_{\alpha=3/2}$-$r(t)$ are in perfect agreement with t-distribution with approximately $\nu_0=3.43\pm 0.05$ and $\nu_0=1.13\pm 0.05$ degrees of freedom, respectively (as we will see later, this difference in the value of $\nu_0$  is unimportant because it can be simply due to the length effects of the finite-time series.). However, the distribution of the log-returns generated by the 2D-L\'evy$_{\alpha=3/2}$-$\ell(t)$ slightly deviates from t-distribution with an obvious difference observable around the zero returns (see Supplementary Fig. S2(c)). Therefore, in what follows we will provide further supporting evidence to prove the compatibility between S\&P prices and the data predicted by 2D-L\'evy$_{\alpha=3/2}$-$r(t)$.


\textbf{Construction of the random matrix analysis.} Complex interactions between different financial assets induce cross-correlations between them and their dynamics which play central roles in the analyses of portfolio management, risk management, investment strategies, etc. The presence of complex interactions, as well as the constant influence of these markets, as a subsystem of their surrounding, from external factors make it difficult and sometimes impossible to accurately estimate the nature of the involved correlations. In addition, the true analysis of real-financial markets suffers from a finite-time evolution which makes the ratio $Q=T/N$ between the length of the financial price time series $T$ and the number of assets $N$ a very relevant parameter in our analysis. 
Although larger ratios would lead to better estimations, but for practical limitations, the
ratio can be even smaller than unity \cite{pharasi2019complex}.

We apply the techniques of RMT to classify the involved correlations in financial markets as complex systems. Our approach provides a framework to simultaneously consider cross-correlations among the assets and the epoch size over which the empirical correlations are developed. 
To start with, we first build matrices of normal returns constructed from day returns of $N=262$ U.S. S\&P stocks. We consider $T\times N$ matrices containing data for $T$ consecutive days as follows 

\begin{figure}[t]
	\centering
	\includegraphics[width=3.2in]{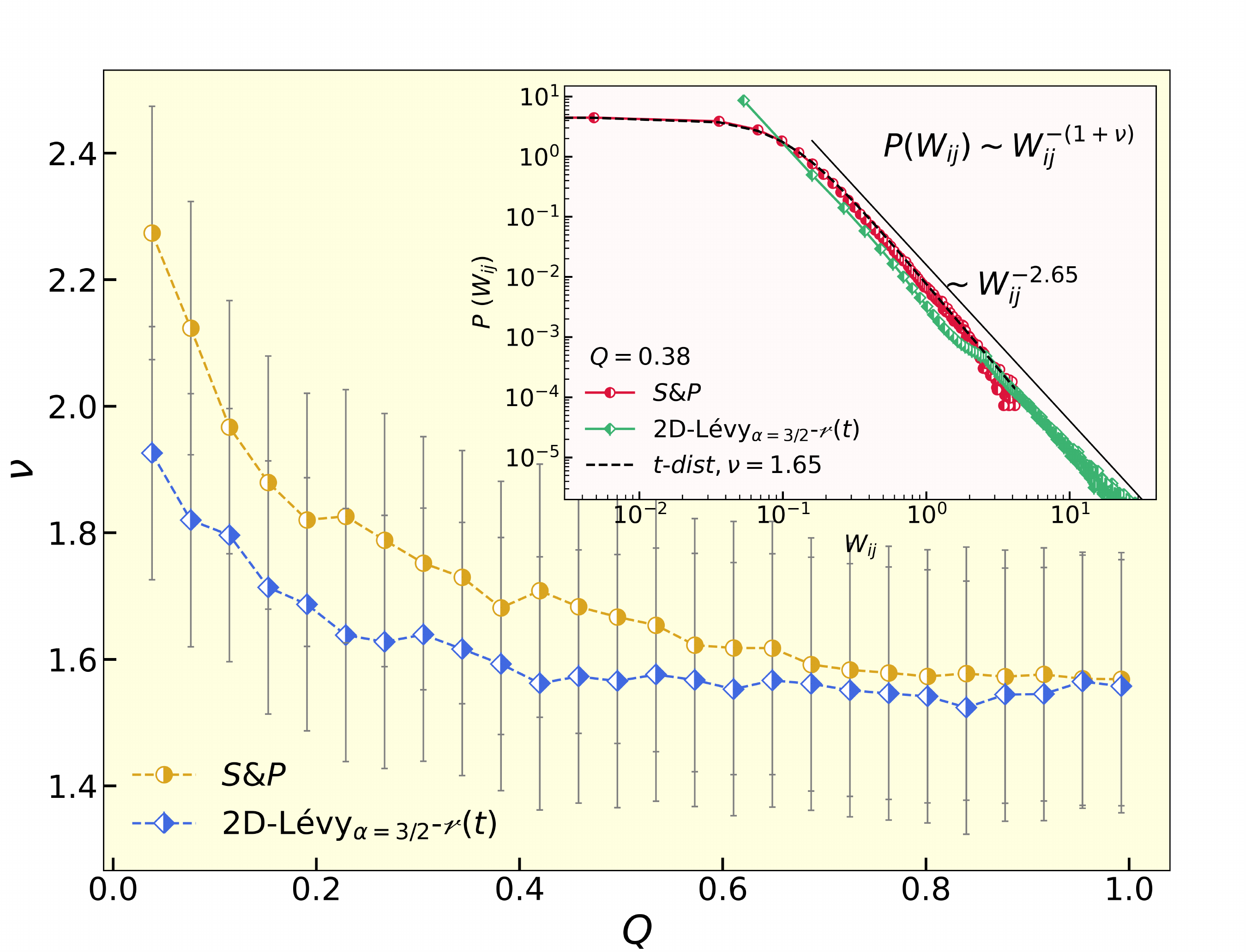}
	\caption{Inset: The distribution of elements of the Wishart matrix with $Q=0.38$ for both the empirical S\&P and 2D-L\'evy$_{\alpha=3/2}$-$r(t)$ model shows overlapping power-law tails. Main: The scaling exponent $\nu$ (see Eq. (\ref{Welements})) as function of $Q$ estimated for both data sets.} 
	\label{fig2}
\end{figure}

\begin{equation}\label{X(t)}
\begin{split}
	&X(t)=\\
	&\begin{pmatrix}
		\delta R_1(t) & \delta R_2(t) & \cdots & \delta R_N(t) \\
		\delta R_1(t+1) & \delta R_2(t+1) & \cdots & \delta R_N(t+1) \\
		\vdots&\vdots&\ddots&\vdots \\
		\delta R_1(t+T-1) & \delta R_2(t+T-1) & \cdots & \delta R_N(t+T-1) \\
	\end{pmatrix},
\end{split}
\end{equation}
whose elements have approximately zero mean and unit variance. Our analysis runs for various epoch size $T=10, 20, 30,..., 260$ to obtain a full quantitative behavior of the system as a function of the ratio $0<Q<1$. The symmetric matrix $W$ is then constructed as follows
\begin{equation}\label{wishart}
	W(t)=\frac{1}{T}X^{\dagger}(t)X(t),
\end{equation}
where $(\cdot)^{\dagger}$ denotes the transpose of the matrix.  $W$ belongs to the type of matrices often referred to as Wishart matrices in multivariate statistics \cite{muirhead1982aspects}. It would be instructive if we first look at the distribution of the Wishart elements constructed in (\ref{wishart}) at different $Q$. This has been shown in the Inset of Fig. \ref{fig2} for $Q=0.38$ as an instance for both S\&P and 2D-L\'evy$_{\alpha=3/2}$-$r(t)$.
For both empirical and model-based data we find a power-law tails in the distribution that share the same exponent $\nu\simeq 1.65$. We find that for the whole range of the ratio $0<Q<1$, the tails of the distribution show a scaling behavior 
\begin{equation}\label{Welements}
P(W_{ij}) \sim W_{ij}^{-(1+\nu)},
\end{equation}   
with an exponent $\nu$ which monotonically decreases from $\nu\sim 2.5$ for small $Q\approx 0$ to $\nu\sim 1.5$ for larger $Q\approx 1$ for S\&P data. The exponents as function of $Q$ are shown in Fig. \ref{fig2} for both S\&P and 2D-L\'evy$_{\alpha=3/2}$-$r(t)$ data sets. For $Q\rightarrow 1$ the exponents for both empirical and model-based data converge, while due to the dominance of fluctuations in short epochs they slightly deviate at small $Q$. 

\begin{figure}[t]
	\centering
	\includegraphics[width=3.3in]{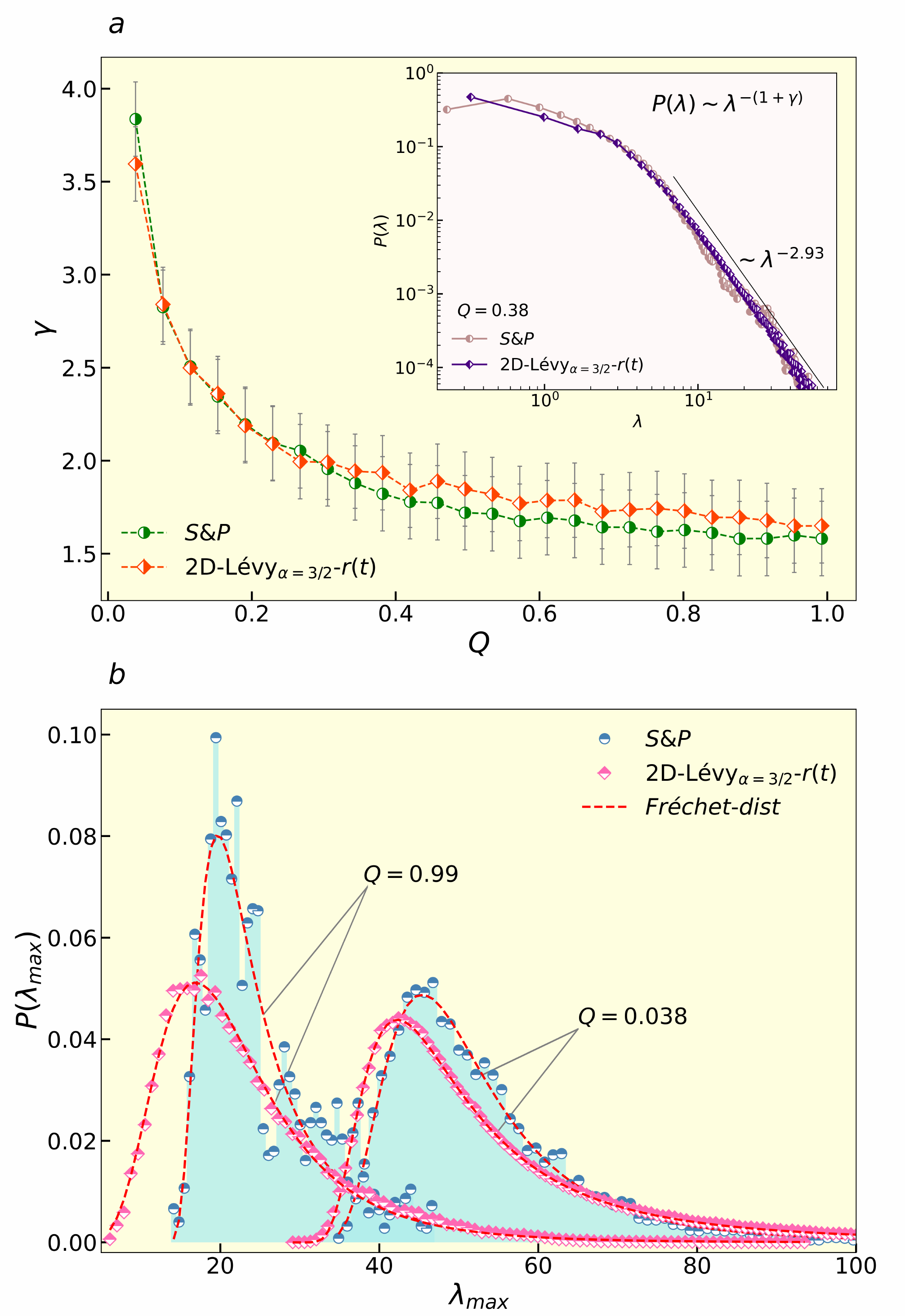}
	\caption{(a) Inset: The distribution of eigenvalues for S\&P and 2D-L\'evy$_{\alpha=3/2}$-$r(t)$ at a given epoch interval $Q=0.38$ collapse on top of each other with a genuine power-law tail of scaling exponent $\gamma\simeq$1.93.  Main: The scaling exponent $\gamma$ as a function of $Q$ are in perfect agreement for S\&P and the model.
	(b) The distribution of the maximum eigenvalues for S\&P and 2D-L\'evy$_{\alpha=3/2}$-$r(t)$ model for $Q=0.038$ and $Q=0.99$ can be well approximated by the Fr\'echet distribution (dashed lines).} 
	\label{fig3}
\end{figure}

\textbf{Epoch-dependent power-law spectra with rare extremes.} The eigenvalue spectrum can be regarded as the fingerprint of the complex networks and be employed to analyze the controllability \cite{goltsev2012localization}, synchronizability \cite{estrada2008communicability}, or the partition of complex networks
into modules or clusters \cite{wang2008vector}.
The statistical properties of a Wishart matrix with uncorrelated elements (a constructed matrix of uncorrelated time series with finite length) are known \cite{bowick1991universal,feinberg1997renormalizing,sengupta1999distributions}. The spectrum of eigenvalues can be evaluated analytically. If there is no correlation between financial indices then the eigenvalues should be bounded between the RMT predictions \cite{kumar2012correlation}. The largest eigenvalues remain stuck at $\sigma ^{2}(1+{Q}^{-1/2})^2$ with Tracy-Widom fluctuations with $\sigma^2$ being the variance of the elements of $X$ (which is $1$ in our case).

\begin{figure}[t]
	\centering
	\includegraphics[width=3.2in]{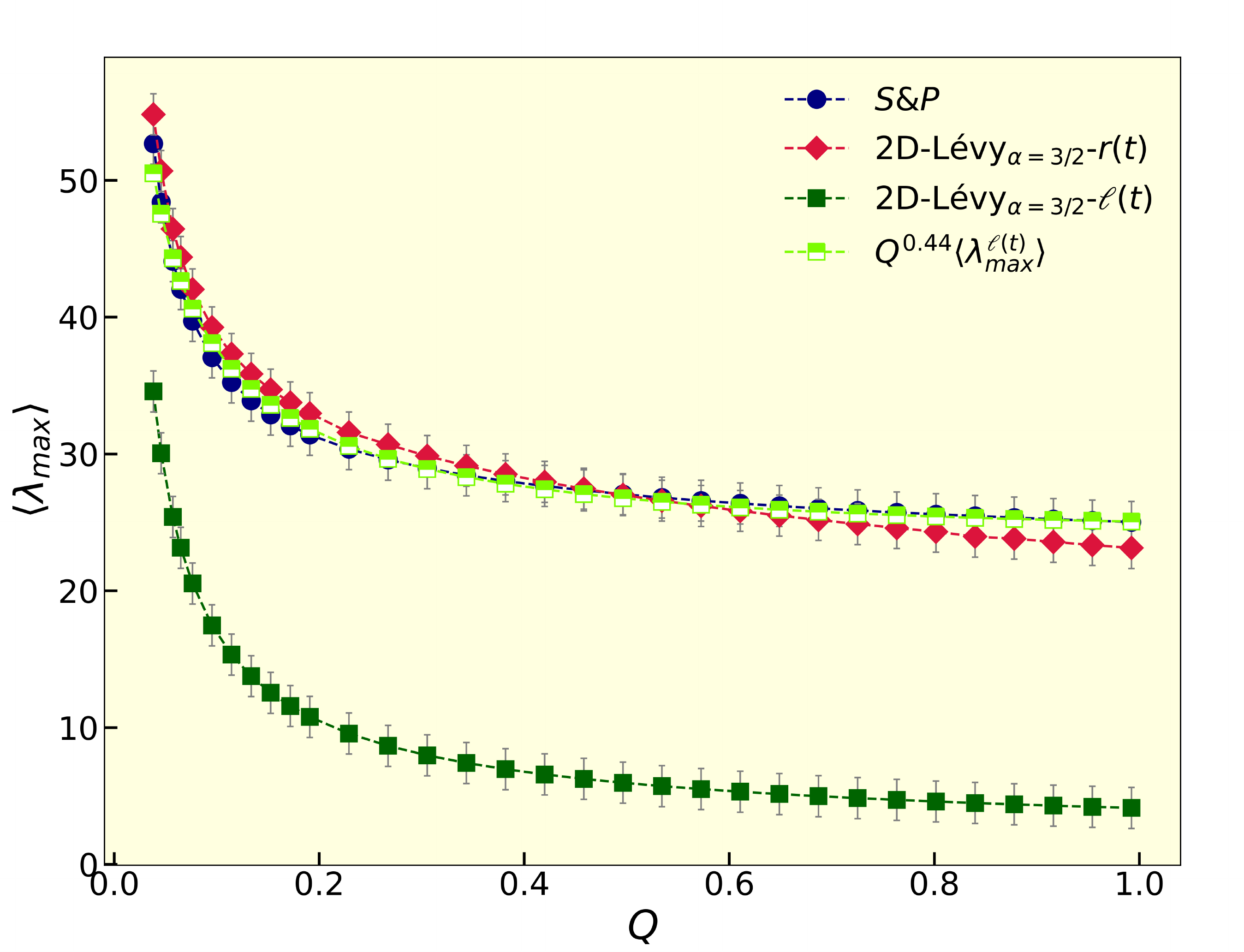}
	\caption{The average maximum eigenvalue as a function of $Q$ for S\&P (solid circles) and 2D-L\'evy$_{\alpha=3/2}$-$r(t)$ model (solid diamonds) collapse on top of each other, but the result $\langle\lambda_{max}^{\ell(t)}\rangle$ for the  2D-L\'evy$_{\alpha=3/2}$-$\ell(t)$ behaves differently (solid squares). The rescaled data $Q^{0.44}\langle\lambda_{max}^{\ell(t)}\rangle$ collapses onto the S\&P result. }.
	\label{fig4}
\end{figure}


Now, we intend to analyze the eigenvalue spectra of an ensemble of $N\times N$ Wishart matrices (\ref{wishart}) constructed for a given epoch length $T$ over the $30$ years of price records. We consider all successive epoch intervals without overlap (we allow for the epoch-overlaps only when we study the statistics of $\lambda_{max}$ for the S\&P markets in order to have enough data for our analysis.).     
For $Q<1$, there exist $N-T$ zero eigenvalues at each sample that we do not consider in our analysis. Thus, each sample gives $T$ nonzero eigenvalues whose distribution is of interest.\\
The Inset of Figure \ref{fig3}(a) shows a sample distribution of nonzero eigenvalues drawn for the S\&P stocks for the epoch with $Q=0.38$. It has a power-law tail for larger eigenvalues with \begin{equation}\label{spectra}
P(\lambda)\sim\lambda^{-(1+\gamma)}
\end{equation} with the exponent $\gamma=1.93\pm0.10$. We have also examined the same analysis for the data generated from our proposed model based on 2D-L\'evy$_{\alpha=3/2}$-$r(t)$ with $Q=0.38$, and we find that the results for both model and S\&P collapse on top of each other by sharing the same tail exponent (see the Inset of Figure \ref{fig3}(a)). We find that the power-law spectra (\ref{spectra}) holds for the whole range of epoch intervals $0<Q<1$ with a monotonically decreasing exponent from $\gamma\sim 4$ for $Q\rightarrow 0$ to $\gamma\sim 1.5$ for $Q\rightarrow 1$ (shown in the Main panel of Figure \ref{fig3}(a)). Most remarkably, our model based on 2D-L\'evy$_{\alpha=3/2}$-$r(t)$ can generate the same observation from S\&P for every epoch interval in the whole range $0<Q<1$ with the same scaling exponent $\gamma(Q)$ at every desired $Q$.

In standard RMT with i.i.d. random real elements of finite variance, the universal limit distribution of the bulk eigenvalues is predicted to be given by the Wigner semicircle law. However, in our case, neither the elements are uncorrelated nor their variance is finite. These cause the observation of power-law distributions for the spectrum of the bulk eigenvalues. The
spectra of the Internet \cite{faloutsos2011power} and scale-free networks
\cite{farkas2011spectra, goh2001spectra, dorogovtsev2003spectra} have been shown to have a power-law tail for large eigenvalues.
 \\ Various properties
of disordered systems and complex networks are sensitive to extreme/edge eigenvalues rather than to typical/bulk eigenvalues. 
Since the eigenvalues of a random matrix are strongly correlated random
variables, one does not expect that the corresponding extremes would belong to any of the three classes predicted by the EVT, i.e, Weibull, Gumbel or Fr\'echet distribution (as discussed earlier). Rather, for a broad class of large Gaussian random matrices, the distribution of the top eigenvalues is predicted to be given by the Tracy–Widom distribution. However, in our study, the power-law distributed eigenvalues especially with the exponent $\gamma<2$ leaves the large eigenvalues uncorrelated and predicts the Fr\'echet distribution for the top eigenvalues (see the Inset of Fig. \ref{fig3}(a)). As shown in Figure \ref{fig3}(b), the distribution of extreme eigenvalues for S\&P stocks at two epoch intervals $Q=0.038$ and $Q=0.99$ are in good agreement with the Fr\'echet distribution. Also, for the top eigenvalues extracted from our model based on 
2D-L\'evy$_{\alpha=3/2}$-$r(t)$ with the same $Q$s, the distributions are in perfect agreement with the Fr\'echet class. One may notice that the deviations between the top eigenvalue distribution of the S\&P and our model is natural since the finite time or size effects are inevitably strong in real-life systems which can lead to effective violations of theoretical predictions.

Despite the discrepancy in the distributions of $\lambda_{max}$ in Fig. \ref{fig3}(b) for a given $Q$ between S\&P and the model, when we look at the average value of the top eigenvalue, for both S\&P and our 2D-L\'evy$_{\alpha=3/2}$-$r(t)$ model, $\langle\lambda_{max}\rangle$ collapse on top of each other within the whole range of the epoch intervals $0<Q<1$ (Figure \ref{fig4}). Once again, this agreement supports the suitability of our proposed 2D-L\'evy$_{\alpha=3/2}$-$r(t)$ to model the dynamics of S\&P stocks.

A similar analysis for 2D-L\'evy$_{\alpha=3/2}$-$\ell(t)$ model leads to completely different results, so that the  distribution of eigenvalues takes on an exponential tail rather than a power-law tail (see the Supplementary Figure S3(a)---please note the semi-logarithmic scale). In addition, the distribution of the maximum eigenvalues are most compatible with the Gumbel universality (Supplementary Figure S3(b)), which is quite different from the observed results for the S\&P. We have also measured the average maximum eigenvalue as a function of $Q$. The results are shown in Fig. \ref{fig4} by solid (dark green) squares. Again, the results are very different from those obtained for the S\&P and 2D-L\'evy$_{\alpha=3/2}$-$r(t)$ model. In order to have a measure for the amount of discrepancy, we note that the rescaled average maximum eigenvalues $Q^{0.44}\langle\lambda_{max}\rangle$ collapse onto the S\&P results. 

\begin{figure}[t]
	\centering
	\includegraphics[width=3.3in]{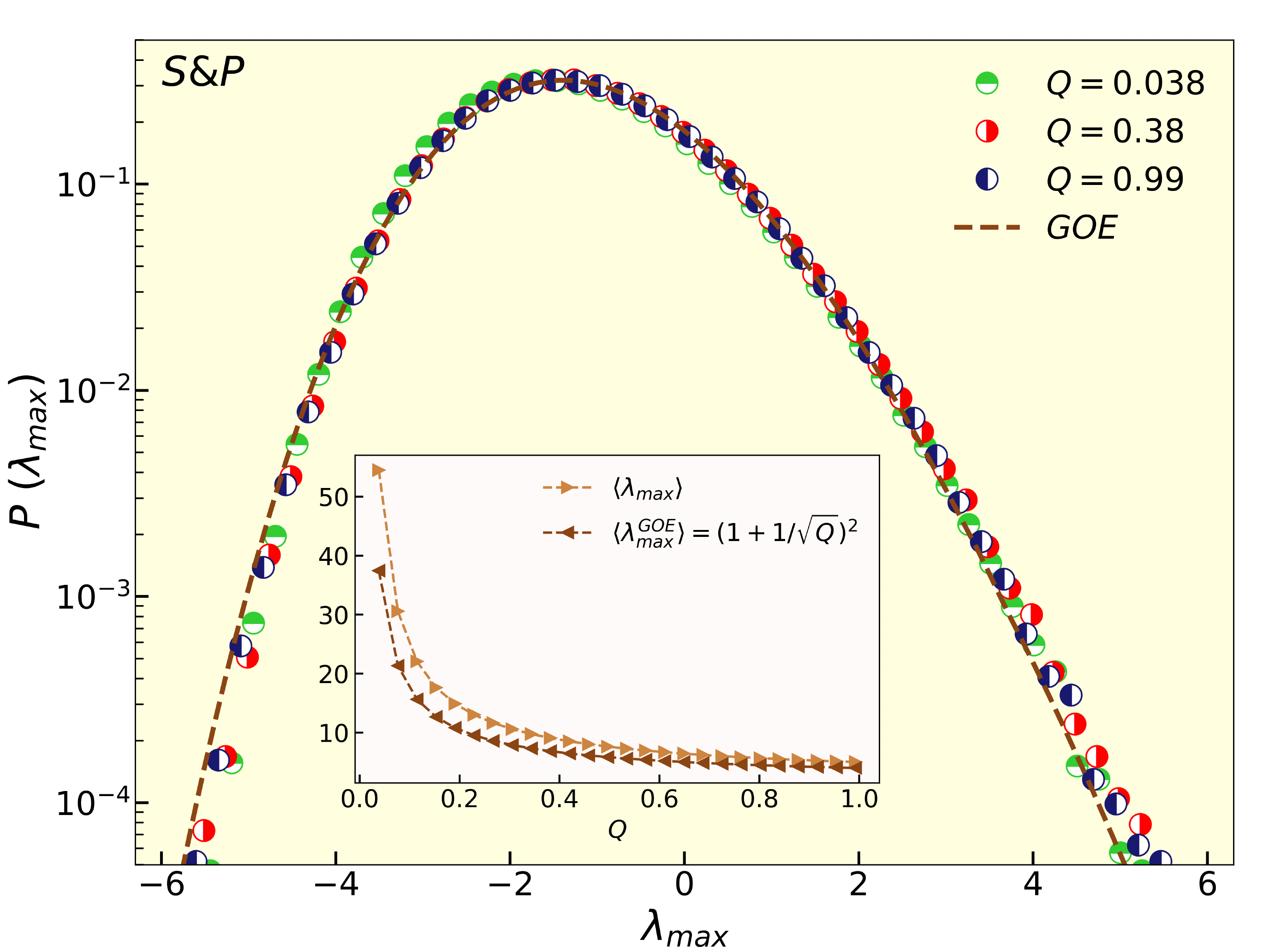}
	\caption{Main: Rescaled distribution function of the maximum eigenvalues obtained for the shuffled S\&P data for three epoch intervals $Q=0.038, 0.38$ and $0.99$. The shuffling procedure destroys the correlations among the data while it leaves their distribution intact. The dashed line corresponds to the well-known GOE Tracy-Widom distribution. Inset: The average maximum eigenvalue as a function of $Q$ compared with the existing theoretical prediction for the GOE ensemble. } 
	\label{fig5}
\end{figure}

\textbf{Randomization.} To examine the importance of correlations in the behavior observed in the analysis of S\&P data, let us perform a similar analysis for its shuffled data. To this end, for a constructed matrix $X(t)$ at any given $Q$ in (\ref{X(t)}), we randomly rearrange the position of all  elements in the matrix and look at the distribution of the maximum eigenvalues of the corresponding Wishart matrices. As Figure \ref{fig5} shows, for all choices of $Q=0.038$, $0.38$ and $0.99$, the rescaled distribution of the maximum eigenvalues agrees well with the GOE Tracy-Widom distribution (the dashed line) known in the standard RMT. We also find a good agreement between $\langle\lambda_{max}\rangle$ and that predicted for the GOE ensemble as a function of $Q$ (see the Inset of Fig. \ref{fig5}).

\textbf{Conclusions.} The main highlight of our present study is the use of L\'evy flights to delicately model the daily closing values of the S\&P 500 index prices. The characteristic features of our devised model are that (i) it attributes a specific spatial dimension $D=2$ and index $\alpha=3/2$ to the proposed L\'evy flights, (ii) it models the consecutive price changes not by the length of the steps but by the difference in the distance from origin, and (iii) it classifies the dynamical behavior of the stock markets within a desired epoch interval $Q$ in terms of a unique scaling exponent obtained from a characteristic power-law eigenvalue spectra. The real-life financial markets are facing with the limited number $N$ of agents and finite recorded time $T$ intervals. However, our 2D-L\'evy$_{\alpha=3/2}$-$r(t)$ model allows for the extrapolations to the $N\rightarrow \infty$ and $T\rightarrow \infty$ limits with exact mathematical predictions. Moreover, it is possible to model the interaction network and topology as well as the clustering properties of the involved stocks based on the similar spectra with the scale-free complex networks. Our study also motivates further research to evaluate the response of financial markets to different kinds of external perturbations and the controbility of real-life networks.  

\section*{Supplementary Material}
Supplementary Material includes the details of our numerical simulations for generation of 2D L\'evy flights and estimation of the fractal dimension of a random trajectory. The distributions of log-returns produced from the S\&P daily prices and the output of our studied models are also presented in the Supplementary Material. It also demonstrates the exponential decay of eigenvalue distribution for the 2D-L\'evy$_{\alpha=3/2}$-$\ell(t)$ at various time intervals whose extremes are shown to be given by the Gumbel distribution. 

\section*{Acknowledgment.} 
 We would like to thank the High Performance Computing (HPC) center in the University of Cologne, Germany,
where a part of computations have been carried out.

\section*{AUTHOR DECLARATIONS}
\subsection*{Conflict of Interest}
The authors have no conflicts to disclose.

\section*{DATA AVAILABILITY}
The data that support the findings of this study are available
from the corresponding author upon request.
		
\bibliography{refs} 

\end{document}